# A Lightweight Deep Learning Model for Human Activity Recognition on Edge Devices

Preeti Agarwal*, Mansaf Alam

*Department of Computer Science, Jamia Millia Islamia, Delhi, 110025, India.*

**Abstract**

Human Activity Recognition (HAR) using wearable and mobile sensors has gained momentum in last few years, in various fields, such as, healthcare, surveillance, education, entertainment. Nowadays, Edge Computing has emerged to reduce communication latency and network traffic. Edge devices are resource constrained devices and cannot support high computation. In literature, various models have been developed for HAR. In recent years, deep learning algorithms have shown high performance in HAR, but these algorithms require lot of computation making them inefficient to be deployed on edge devices. This paper, proposes a Lightweight Deep Learning Model for HAR requiring less computational power, making it suitable to be deployed on edge devices. The performance of proposed model is tested on the participant's six daily activities data. Results show that the proposed model outperforms many of the existing machine learning and deep learning techniques.

## 1. Introduction

In the recent years, smart wearables have become an integral part of human's daily lives [1]. These smart wearables can be used to capture individual's daily activity data. One of the most commonly available sensors is accelerometer in built in smartphones, which can be used to capture tri-axial acceleration of an individual. This tri-axial data can be used to identify various daily activities like walking, jogging, sitting, standing, upstairs, and downstairs. Daily activity data can be used to identify lifestyle, patterns of physical activity of an individual. Any deviation from routine activities can be easily identified by monitoring daily activities, as most of the chronic diseases are caused by lack of physical activity. Besides these, HAR has number of applications in elderly care, surveillance, education, military, tactical scenarios, and rehabilitations [14].

Daily Activities, if captured continuously for longer durations become life logging data [15]. Once these activities are captured then are needed to be stored somewhere for making long time predictions. One of the most promising solution now a days is to store on cloud and number of middlewares are available for integrating sensor data with cloud for analysis [23]. But, transmitting accelerometer signals continuously to cloud may result in overwhelming network traffic and increased latency. Identifying daily activities on edge devices before transmitting it to the cloud can eliminate this problem by reducing communication latency, cost, network traffic, and response time [16].

This paper, proposes a Lightweight Deep Learning model for HAR on edge devices. The contributions of the paper are as follows:

- Firstly, the architecture for proposed Lightweight model is described. This model is developed using Shallow Recurrent Neural Network (RNN) combined with Long Short Term Memory (LSTM) deep learning algorithm.
- Secondly, the model is trained and tested for six HAR activities on resource constrained edge device like Raspberry Pi3, using optimized parameters.
- Thirdly, Experiment is conducted to evaluate efficiency of the proposed model on WISDM dataset [1] containing sensor data of 29 participants performing six daily activities: Jogging, Walking, Standing, Sitting, Upstairs, and Downstairs.
- Fourthly, performance of the model is measured in terms of accuracy, precision, recall, f-measure, and confusion matrix and is compared with certain previously developed models.

The organization of paper is as follows: Section 2 presents background of HAR systems and some of the related works. Section 3 describes proposed Lightweight RNN-LSTM model. Section 4 presents experimentation and evaluation of the model. Section 5 presents result and compares the proposed model with existing models. Conclusion with directions for future work is given in Section 6.

## 2. Background and Related Work

The basic architecture for HAR consists of following phases: Data Collection, Data Pre-processing, Feature Extraction, Activity Classification and Evaluation [17], as shown in fig. 1. Once data is collected through various sensors, it needs to be preprocessed for further use. Pre-processing involves application of data cleaning methods, such as, removal of noise, representation of raw signal, dealing with missing values. Cleaned data is then segmented into windows. Different approaches for window segmentation can be sliding window, event based, and energy window [18]. After pre-processing, feature extraction is done to reduce the set of features for increasing classification algorithm efficiency. Handcrafted or Deep Learning methods can be used to perform this task [19]. Extracted features are classified using classification algorithm and their performance is evaluated using various metrics, such as accuracy, confusion matrix, precision, recall, f-measure [14].

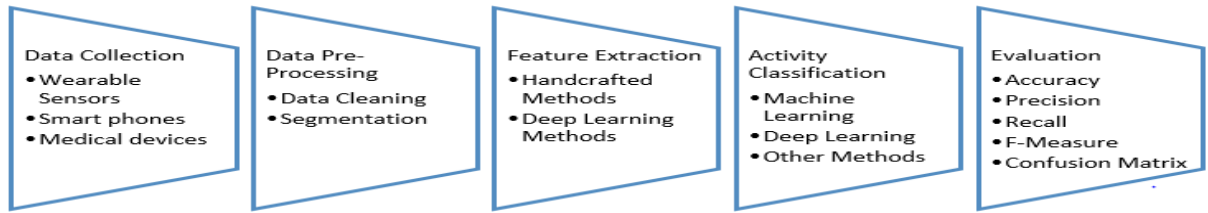

Fig. 1. Basic HAR Steps

Deep Learning methods gained popularity in recent years due to its ability for automatic feature extraction [20]. Many deep learning algorithms have been used for HAR such as CNN ([8], [9], [13], [24]), RNN([10], [12], [21]), Autoencoders [25]. Most of these algorithms suffer from problem of requirement of high computation power, making them difficult to be deployed on edge devices.

Human Activities are encoded as sequence of sensor readings in time T. Traditional machine learning techniques perform classification task without capturing temporal co-relations between input samples. CNN addresses this issue, but is restricted by convolution kernel. RNN combined with LSTM are able to address this issue and have gained popularity for developing HAR systems [21]. In this paper, a Lightweight RNN-LSTM model which can be deployed on edge devices is developed.

## 3. Description of Proposed Lightweight Model

The proposed model is developed using RNN combined with LSTM. It has a shallow structure with just two hidden layers and 30 neurons making it feasible to be deployed on edge computing devices like IoT boards (Raspberry Pi, Audrino, etc.), Android, iOS based resource constrained devices. The description of each component of the model is given below.

### 3.1. Recurrent Neural Networks(RNN)

RNNs are capable of capturing temporal information from sequential data. It consists of input, hidden, and output layer. Hidden layer consist of multiple nodes. The pictorial representation of a typical RNN node is given in fig. 2. Each hidden node t has generating function for generating current value of hidden state $h_t$ and output value $o_t$ as given in equation (1) and (2):

$$h_t = \varepsilon(w_{hh} h_{t-1} + w_{ih} x_t + b_h) \quad (1)$$
$$y_t = \varepsilon(w_{ho} h_t + b_o) \quad (2)$$

where, $x_t$= current input state, $h_{t-1}$= previous hidden state, $h_t$ = current hidden value, $o_t$= current output state, $\varepsilon$ = activation function, $w_{hh}$= weight from hidden to hidden state, $w_{ih}$ = weight from input to hidden state, $w_{ho}$ = weight from hidden to output state, $b_h$ = bias term for hidden state, $b_o$= bias terms for output state. Function $\varepsilon$ is called activation functions such as sigmoid, rectified linear unit, hyperbolic tangent.

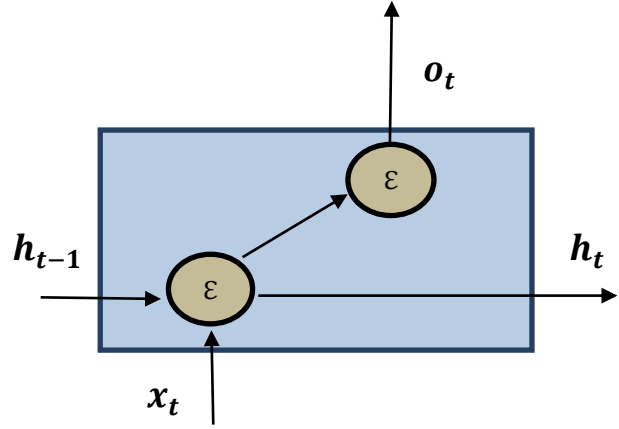

Fig. 2. RNN Node

## 3.2. Long Short Term Memory(LSTM)

RNN networks suffer from the problem of exploding and vanishing gradient. This hinders the ability of network to model wide-range temporal dependencies between input readings and human activities for long context windows. RNNs based on LSTM can eliminate this limitation, and can model long activity windows by replacing traditional RNN nodes with LSTM memory cells. A typical LSTM memory cell is shown in fig. 3. LSTM cell contains various parameters and gates to control behaviour of each memory cell. Each cell state is controlled by the activation functions of gates. The input values are fed to different gates: forget gate (f), input gate (i), output gate (o) and activation vector c. $\varepsilon$ represents activation function.

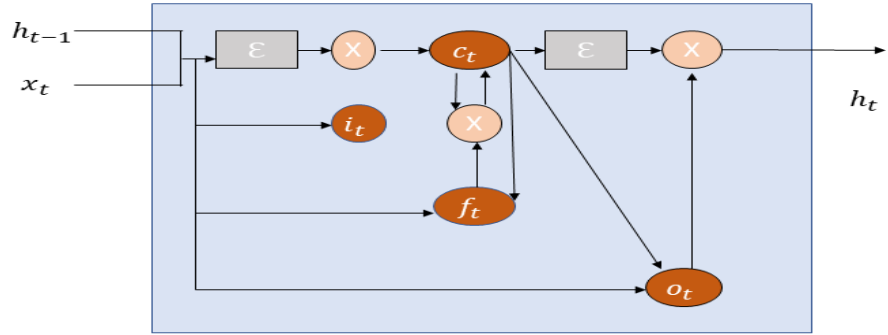

Fig. 3. LSTM Cell

Computation of hidden state value at time step t is according to following equations.

$$i_t = \varepsilon_i(w_{xi}x_t + w_{hi}h_{t-1} + w_{ci}c_{t-1} + bi) \quad (3)$$
$$f_t = \varepsilon_f(w_{xf}x_t + w_{hf}h_{t-1} + w_{cf}c_{t-1} + b_f) \quad (4)$$
$$c_t = f_t c_{t-1} + i_t \varepsilon_c(w_{xc}x_t + w_{hc}h_{t-1} + b_c) \quad (5)$$
$$o_t = \varepsilon_o(w_{xo}x_t + w_{ho}h_{t-1} + w_{co}c_t + b_o) \quad (6)$$
$$h_t = o_t \varepsilon_h(c_t) \quad (7)$$

$w_{xi}, w_{hi}, w_{ci}, w_{xf}, w_{hf}, w_{cf}, w_{xc}, w_{hc}, w_{xo},$ and $w_{co}$, are weights ($w_{xi}$ = input- input weight, $w_{hi}$ = hidden-input weight and so on) and $b_i, b_o, b_c,$ and $b_f$ bias weights.

## 3.3. Proposed Lightweight RNN-LSTM Model

The working of Lightweight RNN-LSTM based HAR system for edge devices is shown in fig. 4. The accelerometer reading is partitioned into fix window size T. The input to the model is a set of readings ($x_1$, $x_2$,

$x_3,\ldots,x_{T-1}, x_T$) captured in time T, where $x_t$ is the reading captured at any time instance t. This segmented window is readings are then fed to Lightweight RNN-LSTM model. The model uses sum of rule and combine output from different states using softmax classifier to one final output of that particular window as $o_T$.

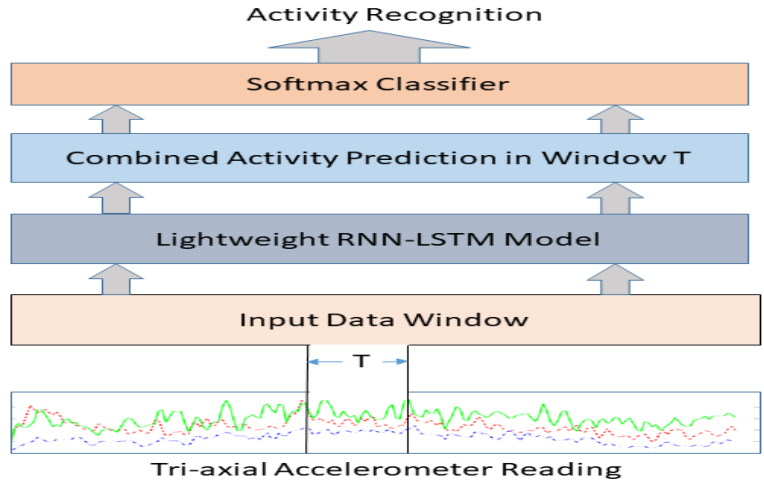

Fig. 4. Working of Lightweight RNN-LSTM Model

The architecture of Lightweight RNN-LSTM model is shown in fig. 5. It consists of two hidden layers with just 30 neurons each. The input $(x_1, x_2, \ldots \ldots, x_T)$ is discrete time signals fed into layer 1 at time $t = \{t = 1,2,\ldots \ldots T\}$. Output of layer one is generated according to the equation

$$o_t^1, h_t^1, c_t^1 = LSTM^1(c_{t-1}^1, h_{t-1}^1, x_t, \theta^1) \quad (8)$$

where $\theta^1$ represents parameters of LSTM cell calculated in equations (3)-(7). Similarly output for the layer 2 can be generated according to the equation (8). The final output is predicted class of activity in window of size T.

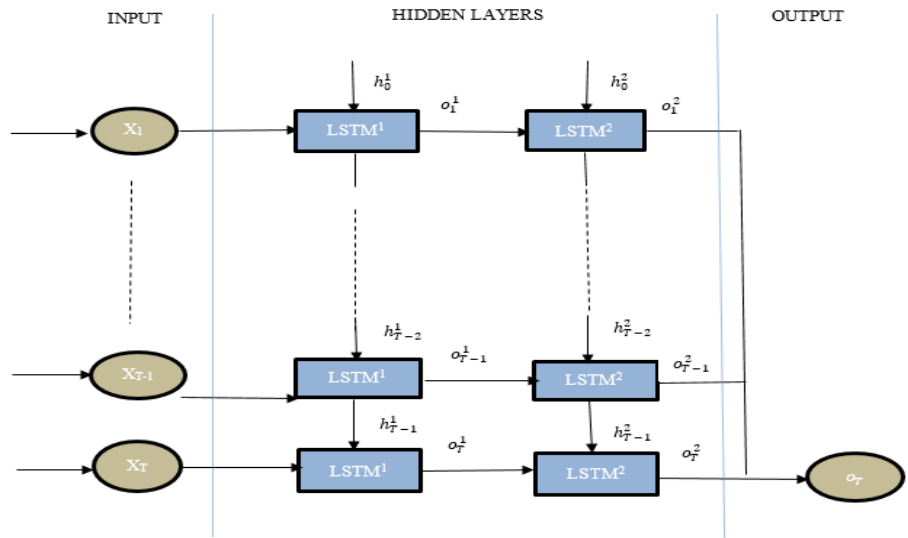

Fig. 5. Architecture of Lightweight RNN-LSTM

## 4. Experiment

The experiment is performed on Raspberry Pi3 with 4xARM Cortex A-53, 1.2 GHz processor, and 1 GB RAM. The model for human activity recognition is implemented on python3.5 and tensorflow1.7. The various steps followed in experimentation is based on basic steps designed by [17].

*4.1. Dataset Description*

Android smartphone having in built accelerometer is used to capture tri-axial data [1]. The dataset consist of six activities performed by 29 subjects. These activities include, walking, upstairs, downstairs, jogging, upstairs, standing, and sitting. Each subject performed different activities carrying cell phone in front leg pocket. Constant Sampling rate of 20 Hz was set for accelerometer sensor. The detailed description of dataset is given in the table 1 below.

Table. 1. Dataset Description

| Total Number of Samples: 1,098,207 | | |
|---|---|---|
| Total Number of Subjects: 29 | | |
| Activity | Samples | Percentage |
| Walking | 4,24,400 | 38.6% |
| Jogging | 3,42,177 | 31.2% |
| Upstairs | 1,22,869 | 11.2% |
| Downstairs | 1,00,427 | 9.1% |
| Sitting | 59,939 | 5.5% |
| Standing | 48,397 | 4.4% |

The acceleration plots for different activities in x, y, z coordinates is shown in fig. 6(a),(b),(c). The plot for jogging shows maximum spikes in x and y direction whereas walking shows maximum spikes in acceleration plot along z direction. Sitting graph has minimum spikes in all x, y and z plots.

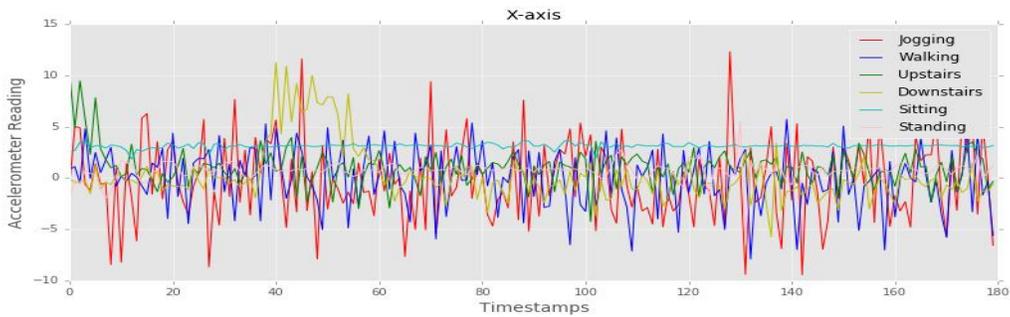

Fig. 6(a) Sensor data plot for activities in X-axis

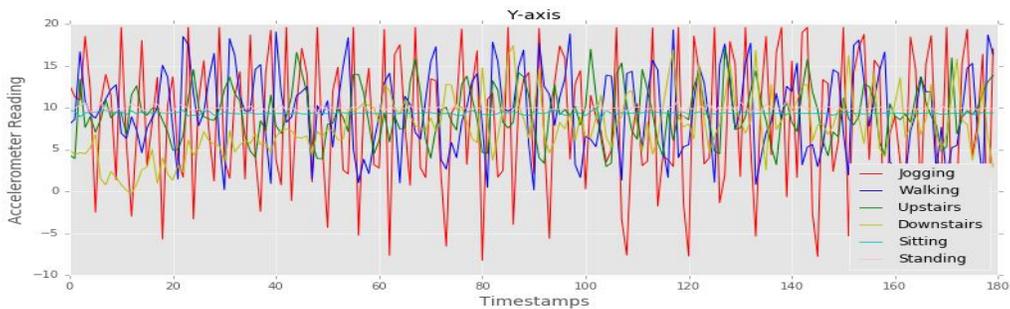

Fig. 6(b) Sensor data plot for activities in Y-axis

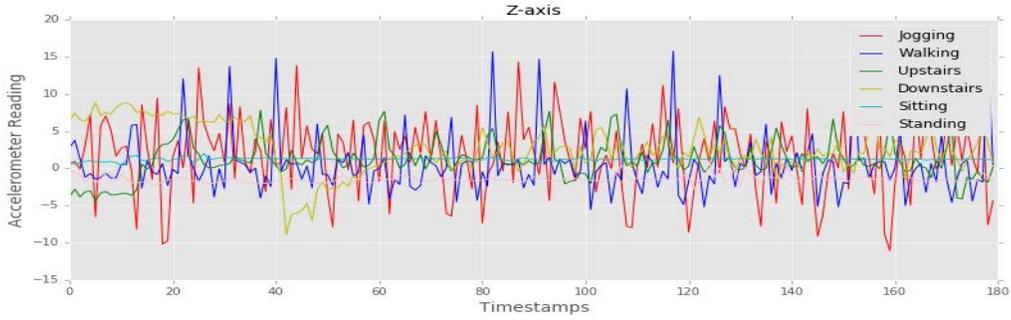

Fig. 6(c) Sensor data plot for activities in Z-axis

*4.2. Training*

Lightweight RNN-LSTM is trained on WISDM dataset split into 70: 30 for training and testing. The weights of the model are updated according to the activation function ε. Mean cross entropy is used as cost function between ground truth and predicted labels. Adam optimizer is used to minimize cost function and update model parameters [24]. This model is trained on Raspberry Pi3 to check the ability of model to work on edge device. Various combinations of parameters such as number of epochs, batch size, window size, learning rate were tried and tested using hit and trial method for Hyperparameters tuning. Final set of selected parameters is listed in table 2. Fig. 7 shows graph between accuracy and training testing cost.

Table 2. List of Selected Hyperparameters

| Category | Hyperparameters | Value Selected |
| --- | --- | --- |
| Data Preprocessing | Time_Step | 100 |
| | Window_Size | 180 |
| | Batch_Size | 64 |
| | Number of Epochs | 75 |
| Architecture | Hidden Layers | 2 |
| | Number of Neurons | 30 |
| Training | Activation Function | Softmax |
| | Bias Weight Initialization | Constant=1.0 |
| Learning | Optimizer | Adam |
| | Learning_Rate | 0.0025 |
| | Loss_Rate | 0.0015 |

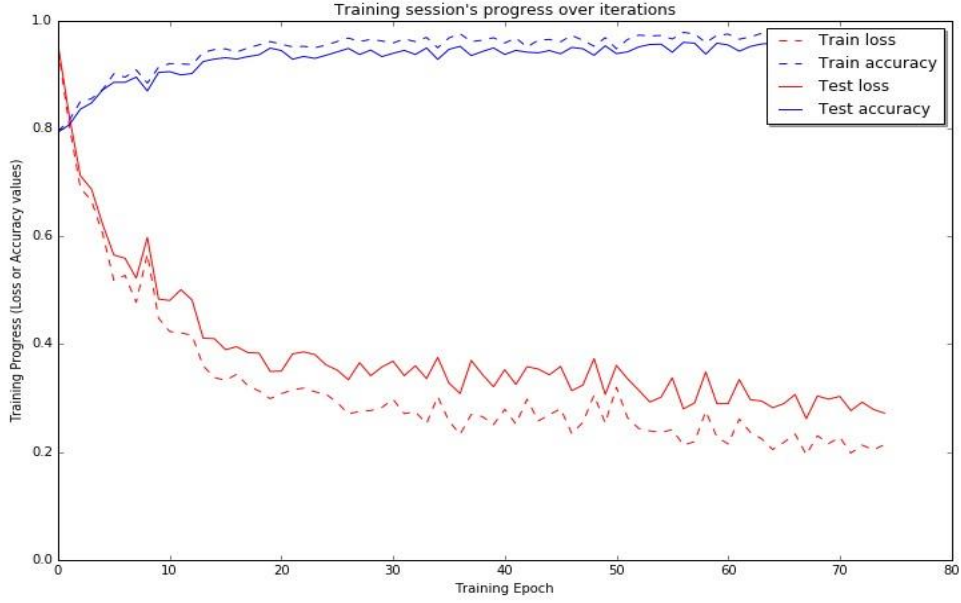

Fig. 7. Training and testing progress over epochs

*4.3. Performance Metrics*

Four widely used evaluation metrics are used to measure the performance of the model namely accuracy, precision, recall and f1-score [22].
- Accuracy: Number of correct predictions over total number of TP predictions made by the model.
$$Accuracy = \frac{TP+TN}{TP+TN+FP+FN} \qquad (9)$$
where, TP = True positive class prediction, TN = True Negative class prediction, FP = False positive class prediction, FN = False negative class prediction.
- Precision: Number of actual true predictions over total true predictions made by the model.
$$Precision = \frac{1}{C}\left(\sum_{c=1}^{C}\frac{tp_c}{tp_c+fp_c}\right) \qquad (10)$$
where, C = total number of classes, $tp_c$ = true positives for a particular class c, $fp_c$ = false positive for a particular class c.
- Recall: number of true predictions over actual number of true predictions made by the model.
$$Recall = \frac{1}{C}\left(\sum_{c=1}^{C}\frac{tp_c}{tp_c+fn_c}\right) \qquad (11)$$
where, C = total number of classes, $tp_c$ = true positives for a particular class c, $fn_c$ = false negatives for a particular class c.
- F1-Score: harmonic mean of the calculated precision and recall.
$$F1 - Score = \sum_{c=1}^{C} 2\left(\frac{n_c}{N}\right) * \frac{precision_c * recall_c}{precision_c + recall_c} \qquad (12)$$
where, N = total number of samples, $n_c$ = number of samples in class c, $precision_c$ = precision value for particular class c, $recall_c$ = recall value for particular class c.

## 5. Result

This section present evaluation results of the Lightweight RNN-LSTM model and compare it with some of the existing works. The confusion matrix of the model is shown in fig. 8. The Lightweight RNN-LSTM achieved 99% accuracy for jogging and walking activity. Minimum accuracy of 81% is achieved for upstairs activity. The results of the evaluation metrics is presented in table 3.

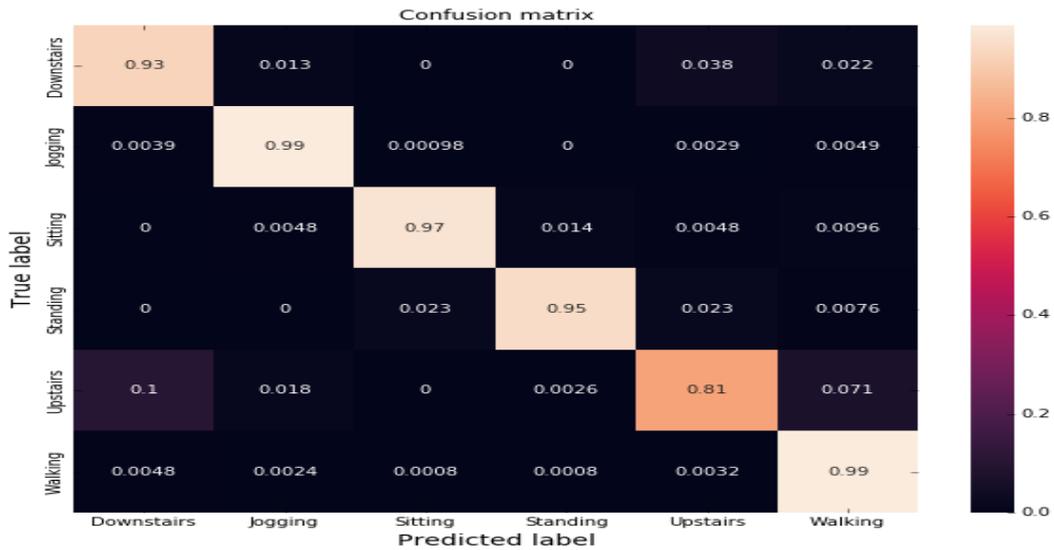

Fig. 8. Confusion Matrix

Accuracy, precision, recall, f1-score is calculated according to equations (9)-(12). Lightweight RNN-LSTM achieved an accuracy of 95.78%, precision value of 95.81%, recall of 95.78%, and F1- score of 95.73%. Accuracy may produce misleading results if the data in each class of dataset is imbalanced, therefore, recall and precision is calculated to validate performance.

Table 3. Evaluation Metrics

|  | Accuracy(%) | Precision(%) | Recall(%) | F1-Score(%) |
|---|---|---|---|---|
| Lightweight RNN-LSTM Model | 95.78 | 95.81 | 95.78 | 95.73 |

This model is compared with some of the previous methods: Deep Learning techniques, such as, CNN and its variations ([13], [9], [8]), RNN([12],[10]), DBN([9]), Machine Learning techniques, such as, SVM([11]), PCA([7]), Hybrid classifiers([6]), Fusion Methods([5]), Hidden Markov Model([4]), Ensemble Learning([2]), Random Forest([3]), and Artificial neural networks, such as Multilayer Perceptron([1]). It outperforms them in terms of accuracy. The graphical representation of comparative results is shown in fig. 9.

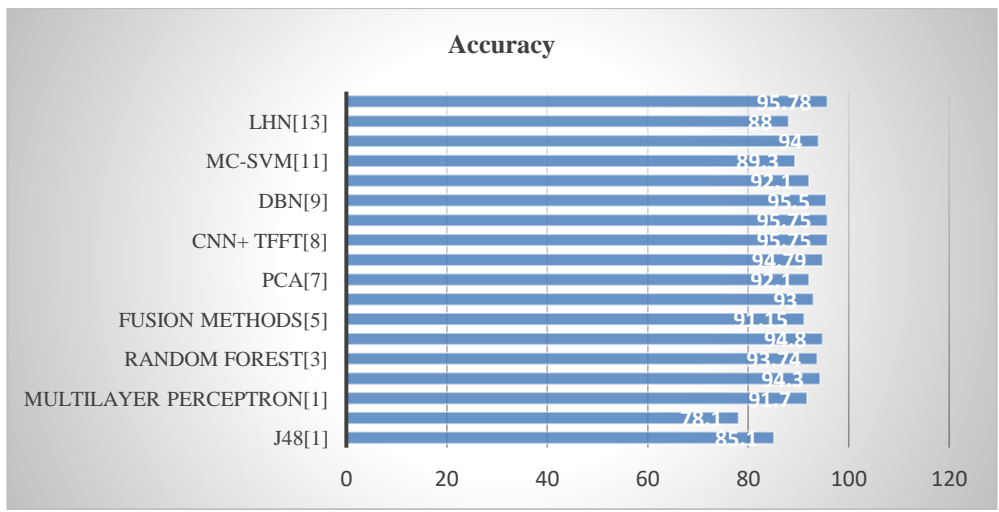

Fig. 9. Comparison of result with existing works

## 6. Conclusions and Future Directions

A Lightweight HAR model is developed in this paper. This model is deployed on edge device Raspberry Pi3. Capturing Human Activities on edge devices reduces communication latency, cost and network traffic. The developed model produces better results as compared to many existing machine learning and deep learning models.

In future, this work can be extended to recognize more complex activities. It can be deployed on other edge devices using Android, iOS. This model is developed using static sliding window architecture. This architecture can also be tested with dynamic windowing scheme in future. Also, this system was developed using single tri-axial accelerometer. It can be extended to support multi-sensor data.

**Acknowledgements**

This work was supported by a grant from ―Young Faculty Research Fellowship‖ under Visvesvaraya PhD Scheme for Electronics and IT, Department of Electronics & Information Technology (DeitY), Ministry of Communications & IT, Government of India.